\documentclass[twocolumn,pra,aps,showpacs,secnumarabic]{revtex4-1}
\usepackage{graphicx}
\usepackage{color}

\usepackage{epsfig}
\usepackage{epstopdf}
\usepackage[T1]{fontenc}
\usepackage{bm}

\def\pH{p-H$_2$}

\newcommand{\beq}{\begin{eqnarray}}
\newcommand{\eeq}{\end{eqnarray}}

\begin{document}
	\setcounter{page}{1}
	
	\title{ Quantum Monte Carlo study of strongly interacting bosonic one-dimensional systems in periodic potentials}
	\author{ K. D\v{z}elalija,  L. Vranje\v{s} 
		Marki\'{c}}
	\affiliation{Faculty of Science, University of Split, Ru\dj era 
		Bo\v{s}kovi\'{c}a 33, HR-21000 Split, Croatia}

	\begin{abstract}
		We present diffusion Monte Carlo (DMC) and path-integral Monte Carlo (PIMC) calculations of a one-dimensional Bose system with realistic interparticle interactions in a periodic external potential.
		Our main aim is to test the predictions of the Luttinger liquid (LL) theory, in particular with respect to the superfluid-Mott insulator transition at both zero and finite temperatures, in the predicted robust and fragile superfluid regimes.
		For that purpose, we present our results of the superfluid fraction $\rho_s/\rho_0$, the one-body density matrix, the two-body correlation functions, and the static structure factor.
		The DMC and PIMC results in the limit of very low temperature for $\rho_s/\rho_0$  agree, but the LL model for scaling $\rho_s/\rho_0$ does not fit the data well. The critical depth of the periodic potential is close to the values obtained for ultracold gases with different models of interaction, but with the same value of the bare LL parameter, demonstrating the universality of LL description. Algebraic decay of correlation functions is observed in the superfluid regime and exponential decay in the Mott-insulator one, as well as in all regimes at finite temperature for distances larger than a characteristic length.
	\end{abstract}


	\maketitle

	\section{Introduction}
	
	Interacting one-dimensional (1D) systems in periodic potentials and disorder present rich phenomena, that have still not been completely explored, despite many years of study. Reviews of 1D systems are given in Refs. \cite{Cazalilla:11,Imambekov:12}. 
	Low-energy phenomena of the system are expected to follow Luttinger liquid (LL) theory \cite{Haldane:81}.  Despite the lack of long-range order, in a uniform system the quasi superfluid phase is predicted  for Luttinger parameter $K>0.5$, while quasi-solid order is expected for $K<0.5$. The superfluidity in one dimension is a finite-size effect, depending on the product of length and temperature, which means it should disappear in the thermodynamic limit for any finite temperature. The correlation functions are predicted to decay algebraically at zero temperature, while at finite temperatures the crossover to exponential decay is expected~\cite{Cazalilla04}.
	
	In periodic potentials the superfluid-Mott insulator transition is predicted and confirmed in experiments with ultracold gases, which are loaded in cigar-like traps and shallow optical lattices~\cite{Haller}.  The transition was explored in the limit of zero temperature theoretically using continuous quantum Monte Carlo simulations, which have  mapped the phase diagram and demonstrated the applicability of the Bose-Hubbard model, which is typically used for deep lattices and  the sine-Gordon model which is often used for shallow optical lattices~\cite{deSoto:12,deSoto:12a,Carbonell-Coronado:13,Carbonell-Coronado:14,Astrakharchik:16,Boeris:16}. Optical lattices were also shown to be a very good tool to study defect-induced superfluidity~\cite{Astrakharchik:17}.
	
	One-dimensional systems can be created also by adsorption in nanopores or nanotubes. Such examples include $^4$He and parahydrogen. He in one dimension was studied in Refs. \cite{Krotscheck:99,Boninsegni:00,Gordillo:00he} and its LL properties were demonstrated by Bertaina \textit{et al.}~\cite{Bertaina:16}.  The LL properties of helium are also observed in quasi-1D environments of  nanopore~\cite{DelMaestro:11,DelMaestro:12,Kulchytskyy:13,nanopore,nanopore1d}. In particular, in Ref. \cite{nanopore} it was shown that 1D-like behavior is only obtained for narrow pores, when two atoms are unable to exchange positions along the axis. In Ref. \cite{nanopore1d}, helium for such nanopores was studied for different densities, demonstrating LL behavior at finite temperature. 
	
	Liquid parahydrogen (\pH) has also been  investigated as a possible 1D superfluid in pure one dimension~\cite{Gordillo:00,Boninsegni:13}, in carbon nanotubes \cite{Gordillo:00,Gordillo:09,DelMaestro:17,Ferre:17}, and in a variety of other nanopores~\cite{Omiyinka:16, boninsengni}.

	Studies of the 1D superfluid-insulator transition in a periodic external potential using continuous models were carried out in the zero-temperature approximation. It is interesting to consider how the interplay of quantum and thermal fluctuations changes the superfluidity and correlations of a strongly interacting system in periodic potentials. Since, according to the LL theory, a system's response  in the low-energy limit should depend on the parameter $K$, we choose a model in which, depending on the density, $K$ assumes values from $\infty$ to 0.  Such behavior in one dimension is demonstrated by the $^4$He atoms~\cite{Bertaina:16} or isotopes of spin-polarized hydrogen~\cite{Vidal:16}.  This allows us to study the regime where superfluidity is expected to be robust ($K$ > 2) and the other one where even an infinitesimal periodic potential is expected to destroy superfluidity~\cite{Cazalilla:11}.
	
	We present the model and methods in Sec. II. The results are presented and discussed in Sec. III. Finally, we summarize our main findings in Sec. IV.

	\section{Model and methods}

	The system under study is composed of $N$ bosons of mass $m$  with the Hamiltonian
	\begin{equation}
	\hat{H}=-\frac{{\hbar}^{2}}{2m}\sum _{i=1}^{N}{\Delta}_{i} + \sum_{i<j}^{N}U(r_{ij}) + \sum _{i=1}^{N}V_{\text{ext}}(x_i),
	\end{equation}
	where $N$ is the number of $^{4}$He atoms of mass $m$,  $r_{ij}=|x_i-x_j|$, $U(r)$ represents the interaction between $^{4}$He atoms modeled  by the Aziz potential~\cite{Aziz:87}, and $V_{\text{ext}}(x)$ is the external potential corresponding to the optical lattice.
	
	In this work, we considered periodic external potential of the form
	\begin{equation}
	V_{\text{ext}}(x)=V_0\sin^2(kx), 
	\end{equation}
	where $k=\pi/a_0$, with $a_0=L/N$ the lattice constant,  so that there is one atom per lattice site. A convenient measure to express the depth is  the recoil energy $E_R=\hbar^2\pi^2/(2ma_0^2)$.

	Finite-temperature calculations were performed using the  worm algorithm path-integral Monte Carlo~\cite{Boninsegni:06,Boninsegni:06a}. The values of the discretized imaginary time $\delta \tau$ were   $8\times 10^{-3}$ K$^{-1}$ for $\rho_0=0.2$~\AA$^{-1}$ and $1.2\times 10^{-2}$ K$^{-1}$ otherwise.
	
	The diffusion Monte Carlo (DMC) method, which solves stochastically the Schr\"odinger equation written in imaginary time was used for zero-temperature calculations, as described in Ref. \cite{Boronat:94a}.
	The DMC method uses a guiding wave function for importance sampling to reduce 
	the variance to a manageable level.   We adopted a Jastrow wave function in the form $\Psi=\prod_{i<j}f(r_{ij})$, where $f(r)=\exp[-(b/r)^5]$. The optimal value of the parameter $b$ was around 3.1 \AA\,. We carefully analysed all possible sources of bias, in particular time-step and population-size bias. Although in most cases the results starting from 2000 walkers were within the errorbars equal to the value obtained by extrapolating to an infinite number of walkers, we typically used 10 000 walkers. The time-step of $\tau$ = 0.3 $\times$ $10^{-4}$ K$^{-1}$ was, in several cases, within the errorbars equal to the value obtained by extrapolation to zero timestep, so we decided to use it for other simulations. 
	
	The superfluid fraction was determined in PIMC calculations using the winding number estimator~\cite{Pollock:87,Ceperley:89}
	\begin{equation}
	\frac{\rho_s}{\rho_0}=\frac{\left\langle W^2\right\rangle}{2\lambda\beta N},
	\end{equation}
	where $\lambda=\hbar^2/2m$, $\beta=(k_BT)^{-1}$, $N$ is the number of particles, and
	\vspace{-3ex}
	\begin{equation}
	W=\sum_{i=1}^{N}\sum_{j=1}^{M}(\textbf{\textit{r}}_{i,j+1}-\textbf{\textit{r}}_{i,j}),
	\end{equation}
	with $M$ the number of time slices.
	In the DMC, the superfluid fraction was determined by an extension of the winding number technique \cite{Zhang:95} which determines the diffusion constant of the center of mass in the limit of infinite simulation time:
	\begin{equation}
	\frac{\rho_s}{\rho_0}=\lim_{\tau\rightarrow \infty} \frac{D(\tau)}{\tau D_0},
	\end{equation}
	where $D_0$ = $\hbar^2/2m$ and 
	\begin{equation}
	D(\tau)=\frac{N}{2}\left\langle \left[x_{\text{c.m.}}(\tau)-x_{\text{c.m.}}(0)  \right]^2 \right\rangle,
	\end{equation}
	with $x_{\text{c.m.}}=\sum_i x_i/N$.
	It is important to track the center of mass of all the particles beyond the simulation cell limits, which means that in this procedure periodic boundary conditions should not be used. 
	
	In our calculations,  for determining superfluidity at each length of the periodic cell and depth of the optical lattice $V_0$, we took an average over ten simulations, using 8000 to 10 000 walkers which were propagated in 30 blocks of 10 000 steps, using the time step 3$\times 10^{-4}$ K$^{-1}$.

	\section{Results}
	
	\subsection{Uniform system}
	We  first studied the properties of the uniform system. At low energies they are expected to follow the Luttinger liquid theory. We determined the equation of state $e=E/N$, which allowed us to calculate the LL parameter as 
	$K=(v_J^0/v_N)^{1/2}=[\pi^2(\frac{\hbar^2}{m})\rho_0^3\kappa]^{1/2}$,~\cite{Cazalilla04} where the compressibility $\kappa$ is given by $\kappa^{-1}=\rho_0\frac{\partial P}{\partial \rho_0}$, with pressure $P = \rho_0^2 \partial e /\partial \rho_0$.
	
\begin{figure}[t]
		\begin{center}
			\includegraphics[width=0.9\linewidth,angle=0]{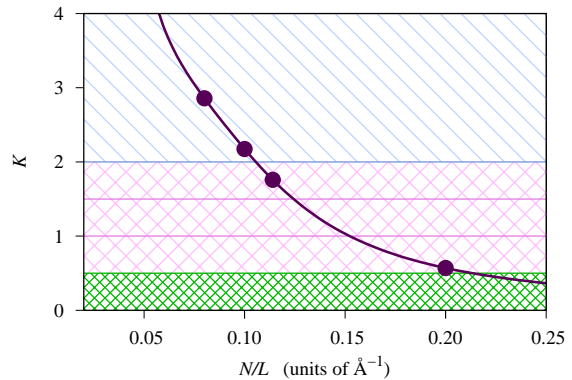}
			\caption{(Color online). Luttinger liquid parameter $K$ as a function of the density.  The bottom area corresponds to the quasi-solid regime with $K<0.5$. The superfluid regime which is robust to periodic potential ($K$ > 2) is at the top. The points at which the response of the system to period potential is studied are marked with circles.}
			\label{fig1}
		\end{center}
	\end{figure}

	The results for the LL parameter, which are in good agreement with Bertaina {\it et al.}~\cite{Bertaina:16}  are given in Fig. \ref{fig1}. Small differences with Ref. \cite{Bertaina:16} are due to different forms of the He-He interaction potential (He-He scattering length here is 88~\AA). In the given density range several physically different regimes are accessible. For $K>0.5$ the system is superfluid, and for $K<0.5$ quasi-solid. The superfluid is predicted to be robust to the periodic potential for $K>2$ and robust to disorder for $K>3/2$~\cite{Cazalilla:11}.
	
	These different regimes are visible in the behavior of  correlation functions at zero and finite temperature. At zero temperature, according to the LL theory, the pair correlation function and the one-body density matrix, at long distances $x\gg a = \rho_0^{-1}$, have the forms
	\begin{eqnarray}
	g(x) &=& 1-\frac{2K}{[2\pi\rho_0 d(x|L)]^{2}}+\sum_{n=1}^\infty\frac{A_n \cos(2\pi n\rho_0 x)}{[\rho_0 d(x|L)]^{2Kn^2}},\label{pdf}\\
	n(x) &=&  \frac{\rho_0}{[\rho_0 d(x|L)]^{1/2K}}\sum_{n=0}^\infty\frac{B_n\cos(2\pi n\rho_0 x)}{[\rho_0 d(x|L)]^{2Kn^2}},\label{obdm}
	\end{eqnarray}
	where $d(x|L)=L|\text{sin}(\pi x/L)|/\pi$ in the case of periodic boundary conditions, and simplifies to $x$ for $L\rightarrow \infty$.
	\begin{figure}[t]
		\begin{center}
			\includegraphics[width=0.9\linewidth,angle=0]{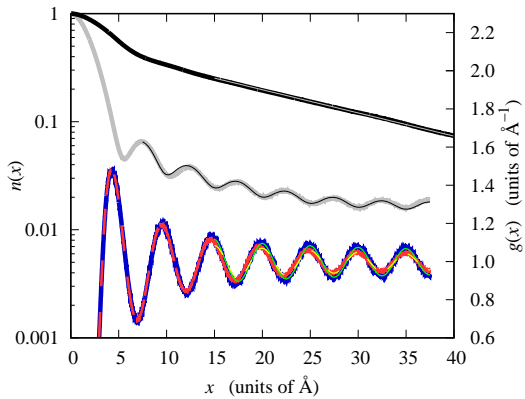}
			\caption{(Color online). Results of the calculations of $n(x)$ and $g(x)$ for uniform system. Upper line represents $n(x)$ for $\rho_0 = 0.1$ \AA$^{-1}$ at $T = 0.12$ K, while middle line is $n(x)$ for $\rho_0 = 0.2$ \AA$^{-1}$ at $T = 0.19$ K. Bottom lines are results of $g(x)$ for $\rho_0 = 0.2 $ \AA$^{-1}$ at $T = 0.04$ K and $T = 0.48$ K, the one with lower peaks (dashed) corresponding to higher temperature. Thin lines are corresponding fits of Eqs. (\ref{pdf}) to (\ref{obdmT}).}
			\label{uni}
		\end{center}
	\end{figure}
	At finite temperature, algebraic decay is observed approximately up to $x \sim L_T$, where $L_T=\hbar v_J/(KT)$, and then it ``crosses over'' to exponential decay~\cite{Cazalilla04}, according to
	\begin{eqnarray}
	g(x) &=& 1-\frac{K}{2\pi^2} \left[ \frac{\pi/L_T}{\rho_0\sinh(\pi x/L_T)}\right]^2  \nonumber \\ 
	&+& B\cos(2\pi\rho_0 x) \left[ \frac{\pi/L_T}{\rho_0\sinh(\pi x/L_T)}\right]^{2K}, \label{pdfT}\\
	n(x) &=& A \rho_0 \left[ \frac{\pi/L_T}{\rho_0\sinh(\pi x/L_T)}\right]^{1/2K}. \label{obdmT}
	\end{eqnarray}
	
	The results of the PIMC calculations of $n(x)$ and $g(x)$ for uniform system are presented in Fig. \ref{uni}. We plot $n(x)$ for $\rho_0$ = 0.2 \AA$^{-1}$ at $T$ = 0.19 K ($L_T$ = 72 \AA) and $\rho_0$ = 0.1 \AA$^{-1}$ at $T$ = 0.12 K ($L_T$ = 14 \AA). In the case of the lower density one observes the exponential decay, while for the higher density one is still effectively in the zero-temperature limit.  In the case of $g(x)$ results are presented for $\rho_0$ = 0.2 \AA$^{-1}$ at two temperatures: $T$ = 0.04 K ($L_T$ = 340 \AA) and $T$ = 0.48 K ($L_T$ = 28 \AA). The effects of exponential decay for lengths larger than $L_T$ are only slightly visible in the decay of the correlation peaks. Overall, the obtained results follow the LL predictions.

	\subsection{Periodic external potential}
	
	Four densities are considered for checking the system's response to the periodic potential, as marked by dots in Fig. \ref{fig1}. Two are in the regime where robust superfluidity is expected and two in the fragile superfluid regime.
	
\begin{figure}
		\centering
		\includegraphics[width=8cm,keepaspectratio=true]{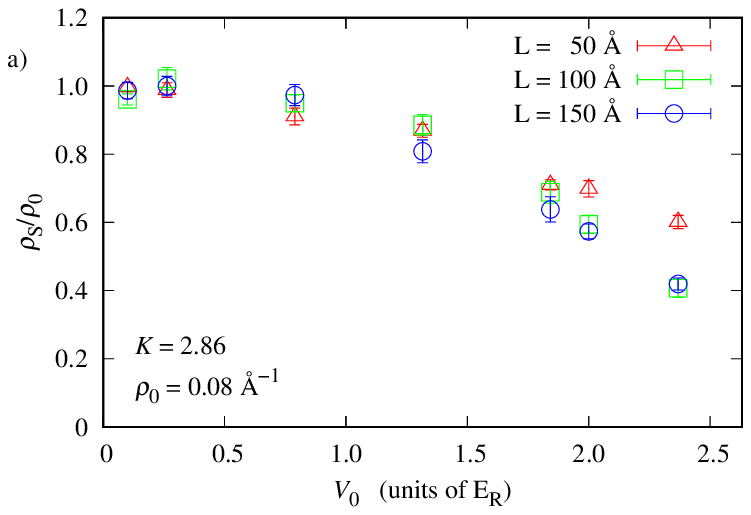}
		\includegraphics[width=8cm,keepaspectratio=true]{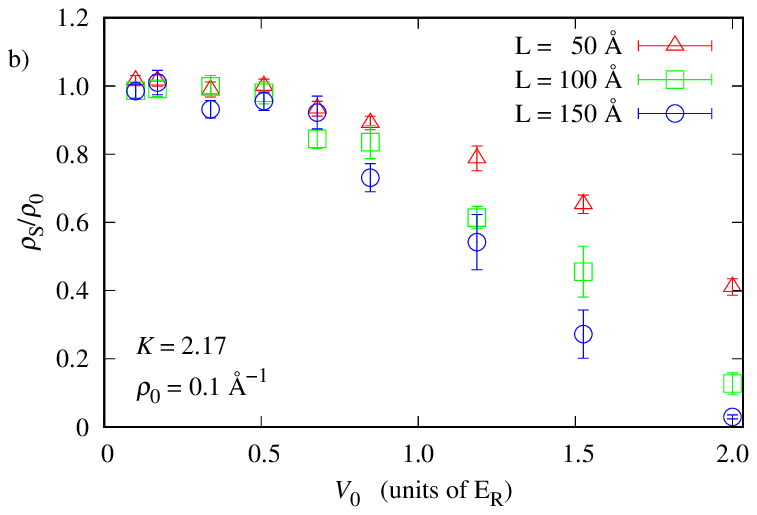}
		\includegraphics[width=8cm,keepaspectratio=true]{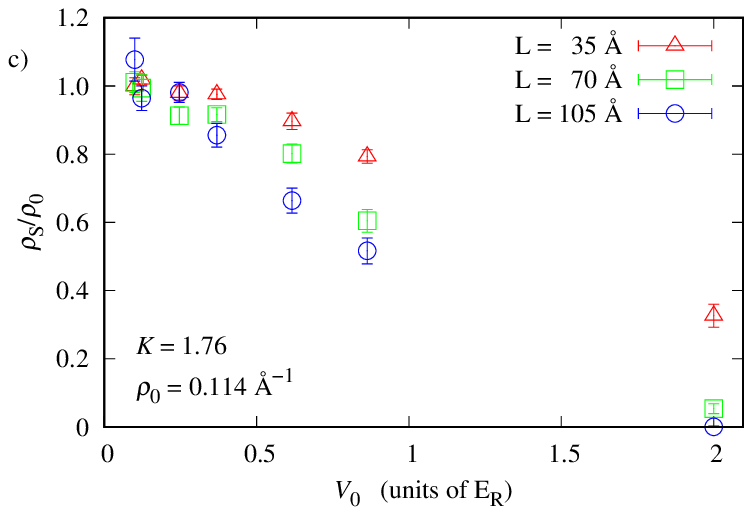}
		\includegraphics[width=8cm,keepaspectratio=true]{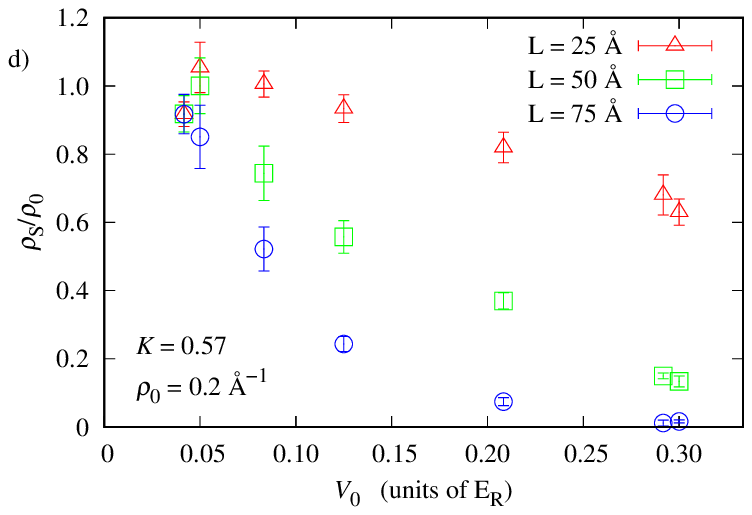}
		\caption{(Color online) DMC results for superfluid fraction at zero temperature for densities: (a) $\rho_0$ = 0.08 \AA$^{-1}$, (b) $\rho_0$ = 0.1 \AA$^{-1}$, (c) $\rho_0$ = 0.114 \AA$^{-1}$, and
			(d) $\rho_0$ = 0.2 \AA$^{-1}$.}
		\label{sfd}
	\end{figure}

	First, in Fig. \ref{sfd}, we present the DMC results for superfluidity at zero temperature as a function of the periodic potential depth $V_0$ in units of the recoil energy ($E_R$), for several lengths of the periodic boundary cell.
	In the superfluid regime, in the limit of zero temperature, the values obtained for different lengths should be the same within errorbars. Thus, when  finite-size superfluidity starts to be observed, that signals the transition to  Mott insulator. 
	For all considered densities and lengths the trend of reducing the superfluid fraction with the depth is observed, but superfluidity persists until some critical depth for $K>2$, as expected.
	Further, the present results show that for larger $K$ one needs larger $V_0/E_R$. Namely, for $\rho_0$ = 0.08 \AA$^{-1}$ the signal for the transition appears between 1.7 and 2 $V_0/E_R$, while for $\rho_0$ = 0.1 \AA$^{-1}$ it is between 0.6 and 0.8 $V_0/E_R$.  However, for the density $\rho_0$ = 0.08 \AA$^{-1}$ the superfluid fraction for two longer lengths overlaps, so we do not have sufficient data to conclude that the superfluid fraction would converge to zero in the limit of infinite length, as expected for Mott insulator.
	For very low potential depths $V_0$ a superfluid fraction appears to saturate very close to 1. It has been shown by Leggett \cite{leggett} that if translation invariance is broken the superfluid fraction, even at $T$ $=$ 0, has to be less than 1. Due to small depths ($V_0 \le 0.1 E_R$) this effect, although expected, is not visible within statistical errors of these calculations.
	It is interesting to compare the results for $K>2$ with experiments on ultracold gases~\cite{Haller,Boeris:16}, in which the critical depth is presented in terms of the Lieb-Liniger parameter $\gamma$~\cite{LiebLiniger}. Since the microscopic model of interaction between the particles differ, we can use the Luttinger parameter $K$ to compare the systems' response. For our density $\rho_0$ = 0.08 \AA$^{-1}$,  $K$ = 2.86, and the same value of $K$ is obtained in Ref. \cite{Haller} for $\gamma$ = 1.5. Different measurements estimate the critical depth $(V_0)_C$  between  1.5 and 2.9 $E_R$, while for the same $\gamma$ in Ref. \cite{Boeris:16} $(V_0)_C$  between 2.4 and 3 $E_R$ is obtained. For the density $\rho_0$ = 0.1 \AA$^{-1}$,  $K$ = 2.17 and the corresponding $\gamma$ = 2.86, for which in Ref. \cite{Haller} $(V_0)_C$ appears at about 0.5 $E_R$, while for Ref. \cite{Boeris:16} it is between 1.2 and 1.4 $E_R$. 
	
	In the case of two higher densities, where  $K<2$, even longer simulations would be needed to determine if the superfluidity disappears at the lowest depths considered. However, it is clear that superfluidity vanishes for $V_0 > 0.1$ $E_R$ in the case of $\rho_0$ = 0.114 \AA$^{-1}$ and $V_0 > 0.04$ $E_R$ for $\rho_0$ = 0.2 \AA$^{-1}$, which is expected from both theory and experiment.  
	
	At finite temperature, according to the LL theory, the superfluid fraction should scale with $LT$. Furthermore, the dynamic superfluid fraction $\rho^D_S/\rho_0$ was introduced~\cite{nanopore1d,Machta:88,Prokofev:00}
	\begin{equation}
	\frac{\rho_S}{\rho_0}=\left(\frac{\alpha_0}{4}\right)\frac{|\Theta_3^{\prime\prime}(0,e^{-\alpha^D/2})|}{\Theta_3(0,e^{-\alpha^D/2})}\label{ds},
	\end{equation}
	where 
	\begin{equation}\label{aD}
	\alpha^D = \alpha_0\left(\frac{\rho^D_S}{\rho_0}\right)^{-1},
	\end{equation}
	$\alpha_0 \equiv \left(TL/\sigma\rho_0\right)$, $\sigma=\hbar^2/k_Bm=12.1193$ K \AA$^2$ and $\Theta_3 (z,q)$ is the Theta function, $\Theta_3^{\prime\prime}(z,q) = d^2\Theta_3 (z,q)/dz^2$. The value of $\rho^D_S/\rho_0$ obtained by fitting the winding number results at different temperatures and lengths should thus correspond to the superfluid fraction results obtained in the zero-temperature DMC calculations in the superfluid regime. In the Mott-insulator regime one does not expect Eq. (\ref{ds}) to be valid, that is, the results for different lengths are not expected to follow the same lines.
	\begin{figure}
		\centering
		\includegraphics[width=8cm,keepaspectratio=true]{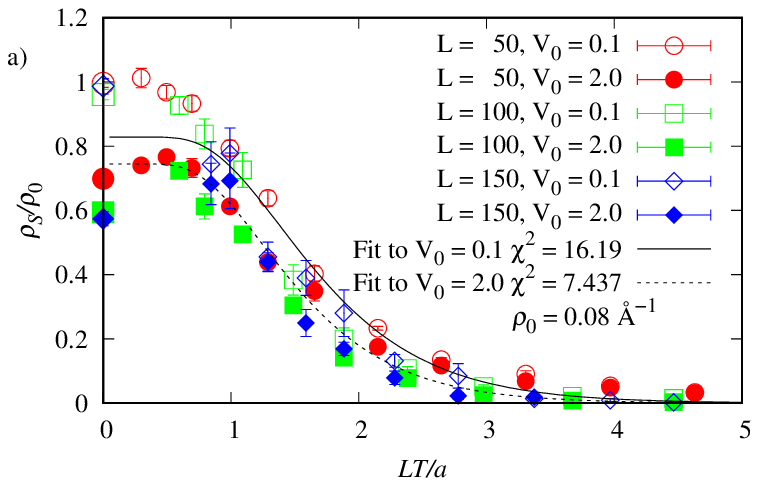}
		\includegraphics[width=8cm,keepaspectratio=true]{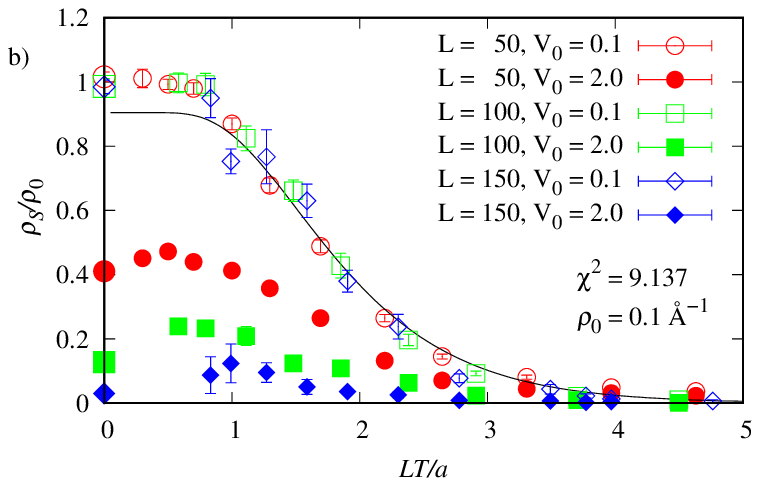}
		\includegraphics[width=8cm,keepaspectratio=true]{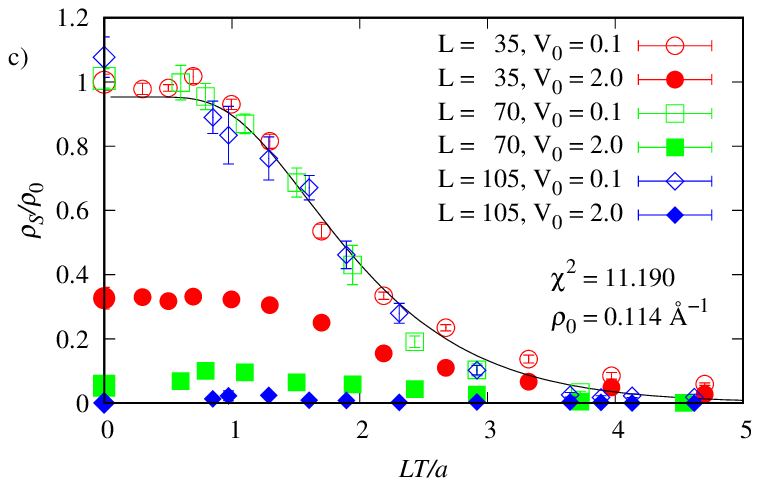}
		\includegraphics[width=8cm,keepaspectratio=true]{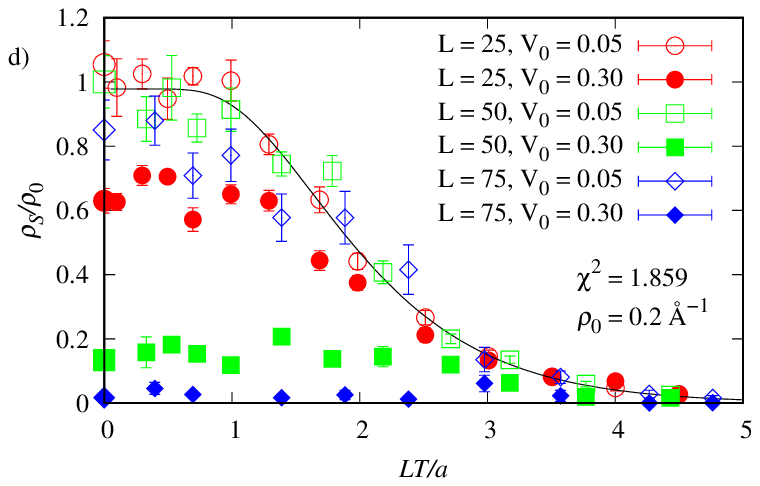}
		\caption{(Color online) Superfluid fraction versus $LT/a$, with $a=\pi(\hbar^2/m)\rho_0$, for densities: (a) $\rho_0$ = 0.08 \AA$^{-1}$, (b) $\rho_0$ = 0.1 \AA$^{-1}$, (c) $\rho_0$ = 0.114, \AA$^{-1}$	and	(d) $\rho_0$ = 0.2 \AA$^{-1}$. Lengths are given in \AA , $T$ in K, depths ($V_o$) in $E_R$. Lines are theoretical fits of Eq. (\ref{ds}) which are presented only when $\chi^2$ is less then 50. We add the DMC results for $T=$ 0 K. }
		\label{scal}
	\end{figure}
	We present in Fig. \ref{scal} the results obtained at different temperatures and lengths for four densities and two depths. Additionally, for $T=0$ K we add the DMC results for comparison. According to the latter, for the first depth  $V_0$ = 0.1 $E_R$ the systems with $\rho_0$ = 0.08 \AA$^{-1}$ and 0.1 \AA$^{-1}$ should be superfluid and for $V_0$ = 2 $E_R$ the Mott insulator is expected in the case of $\rho_0$ =  0.1 \AA$^{-1}$.  Generally, the PIMC results confirm the DMC predictions. Superfluidity appears robust (results for different lengths generally follow the same line) and at low temperatures it is consistent with zero-temperature results in the predicted superfluid regime, obtained by DMC. 

	However,  the fits of Eq. (\ref{ds}) (to data with different lengths and the same value of $V_0$) estimate lower values of $\rho_S/\rho_0$ at zero temperature than DMC results and do not follow the points at low temperatures. This is also reflected in the large $\chi^2$ values. So, it appears that, at low temperatures, the model in Eq. (\ref{ds}), which does not include the external periodic potential, is not completely appropriate.  This is most visible at the lowest depth for $\rho_0$ = 0.08 \AA$^{-1}$.
	
	For $\rho_0$ = 0.08 \AA$^{-1}$ and $V_0$ = 2 $E_R$ there is separation between the results for the smallest length and those at larger ones, similar to the DMC results. However, since the results for two larger lengths overlap within the errorbars, one can not definitely conclude that the SF-Mott insulator barrier has been crossed. For the larger density $\rho_0$ = 0.1 \AA$^{-1}$ and $V_0$ = 2 $E_R$, the results at all temperatures indicate convergence to zero with the increase of the length, as expected for the Mott insulator.

	For $\rho_0$ = 0.114 \AA$^{-1}$, at the lowest depth of 0.1 $E_R$ the systems within the errorbars appear superfluid (just like in DMC).  
	At the higher depth presented, the results for different lengths start to deviate, the model does not fit the data, while the superfluid fraction for the largest length and every $T$ is essentially consistent with zero, which  indicates a Mott insulator.  Similar behavior is obtained for $\rho_0$ = 0.2 \AA$^{-1}$.
	
	One interesting effect is observed in the case of the Mott-insulator phase. The obtained finite-size superfluid fraction for a particular length in several cases increases with temperature and then starts to decrease again.
	\begin{figure}
		\centering	
		\includegraphics[width=8cm,keepaspectratio=true]{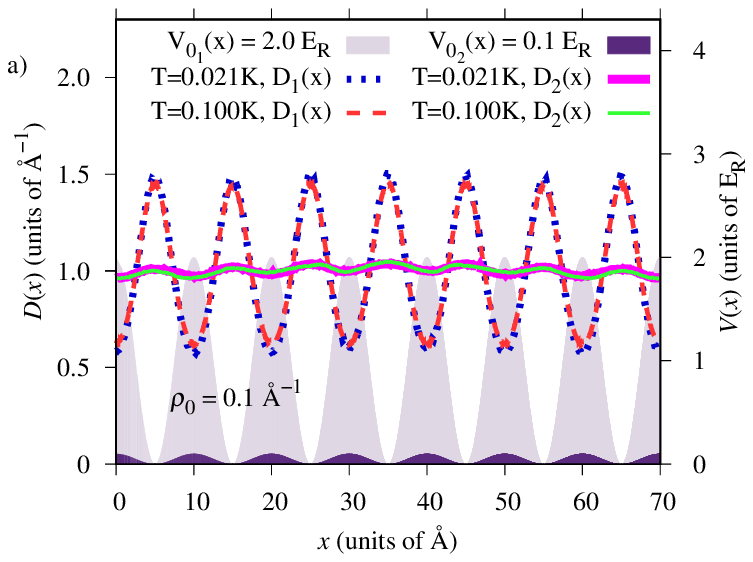}

		\includegraphics[width=8cm,keepaspectratio=true]{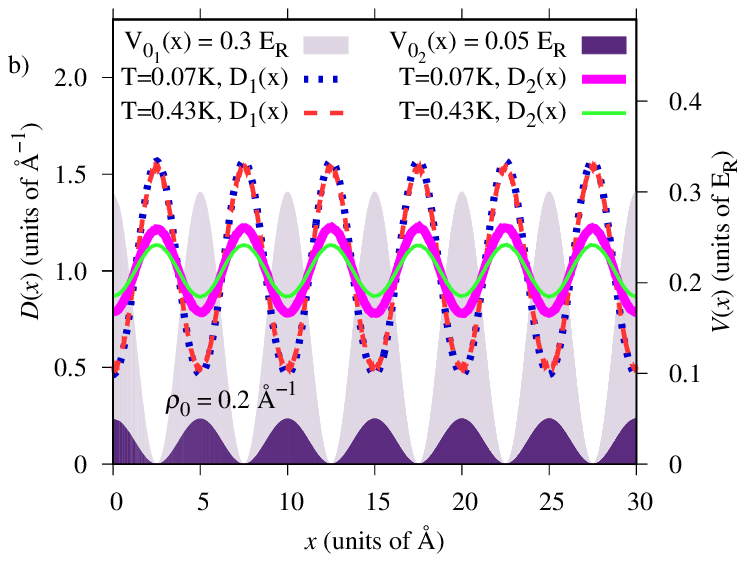}
		\caption{(Color online) Density profile $D(x)$ and external potential $V(x)$ for densities  (a) $\rho_0$ = 0.1 \AA$^{-1}$ and (b) $\rho_0$ = 0.2 \AA$^{-1}$.}
		\label{dens}
	\end{figure}
	
	To gain a better understanding of the system we plot the external potential and the density along the axis for $\rho_0 = 0.1$ \AA$^{-1}$ [Fig. \ref{dens}(a)] and $\rho_0 = 0.2$ \AA$^{-1}$ [Fig. \ref{dens}(b)] for the same configurations as in the Figs. \ref{scal}(b) and \ref{scal}(d).

	One can observe that for the lower $V_0$ and even for $\rho_0$ = 0.2 \AA$^{-1}$ the density profile is only slightly modified. Two temperatures are chosen, one representing the lower and one the higher range on Fig. \ref{scal}.  The temperature increase slightly flattens the periodic oscillations in the density along the axis.  When the superfluidity disappears, the particles are still not completely localized. It does not happen even for the highest density considered  $\rho_0$ = 0.2 \AA$^{-1}$. To achieve complete localization the depth needs to be increased considerably, e.g., to 25 $E_R$ for the density $\rho_0$ = 0.114 \AA$^{-1}$. 
	
	\begin{figure}
		\centering
		\includegraphics[width=8cm,keepaspectratio=true]{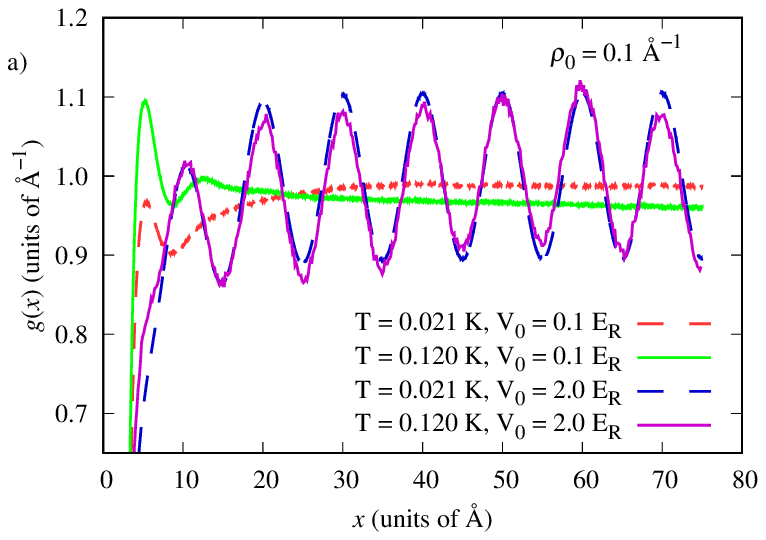}
		\includegraphics[width=8cm,keepaspectratio=true]{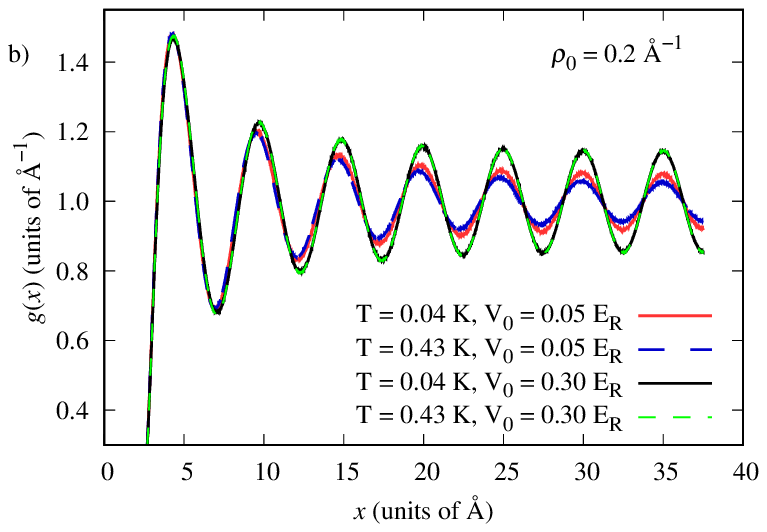}
		\caption{(Color online) Finite-temperature pair-correlation function for two densities (a) $\rho_0$ = 0.1 \AA$^{-1}$ and (b)  $\rho_0$ = 0.2 \AA$^{-1}$. 	}
		\label{gr}
	\end{figure}
	We also calculated the pair correlation functions, which are presented at two densities and two depths in Fig. \ref{gr}. When the lower depth $\rho_0$ = 0.1 \AA$^{-1}$ is considered there is almost no difference with respect to the case without optical lattice. With the increase of external potential depth, oscillations corresponding to the periodic potential appear. In the case of the density $\rho_0$ = 0.2 \AA$^{-1}$, we only observe the enhancement of oscillations which are already present without the optical lattice because the system is near the quasi-solid regime. Additionally, for $\rho_0$ = 0.2 \AA$^{-1}$ the temperature effect is not visible for higher depths, that is, all lines coincide. 
	
	We further calculated the one-body density matrix. The results are presented in Fig. \ref{obdm}. Again, the results are very similar to the case without the optical lattice when the depth is small. In the robust superfluid regime [Fig. \ref{obdm}(a)], for $V_0 = 0.1$ $E_R$ and at very low temperature $T = 0.021$ K algebraic decay is observed as expected, because $L_T$ = 80 \AA. It  crosses over to exponential eventually, which is clearly observed at a higher temperature of $T = 0.12$ K, where $L_T = 14$ \AA. Interestingly, even in the case of fragile superfluid $\rho_0$ = 0.2 \AA $^{-1}$ [Fig. \ref{obdm}(a)] the decay is not exponential at the lowest temperature.  For higher depths, where the system is clearly in the Mott-insulator regime, the decay is exponential at all temperatures.   Temperature effects at higher $V_0$ are small, which is consistent with the behavior of other quantities. 
	\begin{figure}
		\centering	
		\includegraphics[width=8cm,keepaspectratio=true]{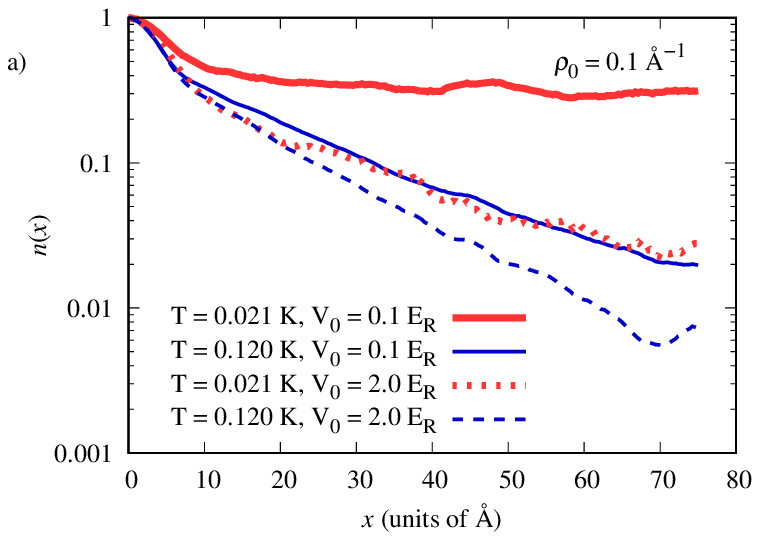}	\includegraphics[width=8cm,keepaspectratio=true]{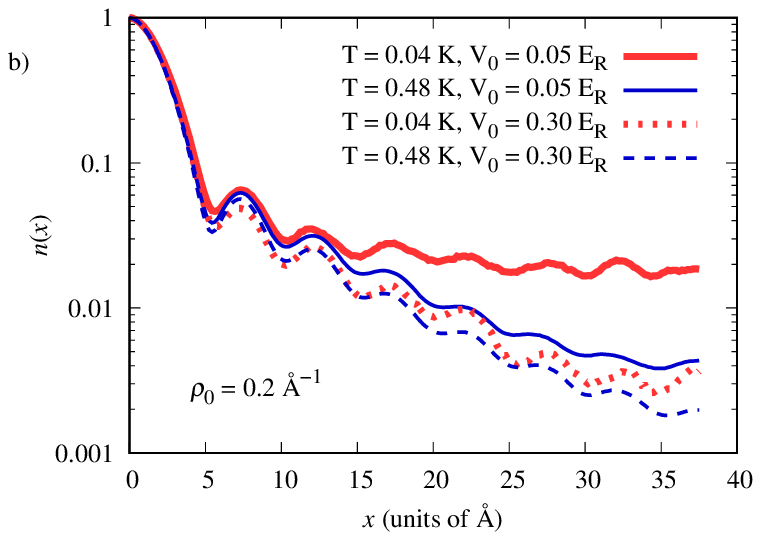}	
		\caption{(Color online) Finite-temperature OBDM for two densities (a) $\rho_0$ = 0.1 \AA$^{-1}$ and (b) $\rho_0$ = 0.2 \AA$^{-1}$ for different temperatures and depths. Statistical errorbars are of the order of the linewidth.}
		\label{obdm}
	\end{figure}
	\begin{figure}
		\centering	
		\includegraphics[width=8cm,keepaspectratio=true]{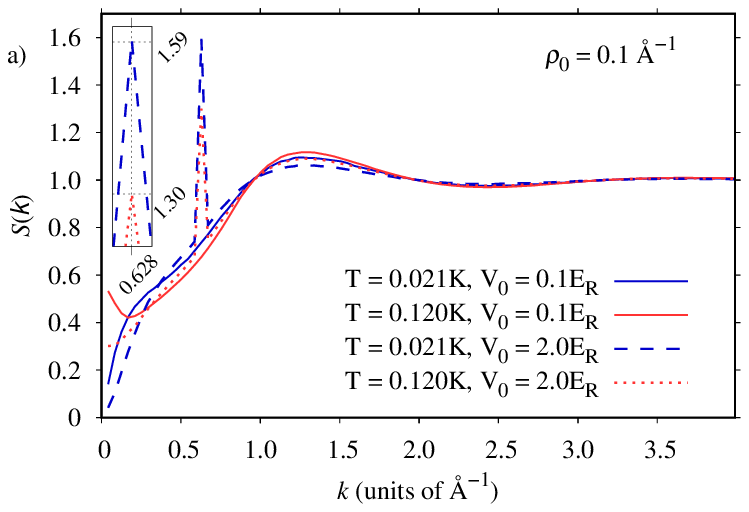}	\includegraphics[width=8cm,keepaspectratio=true]{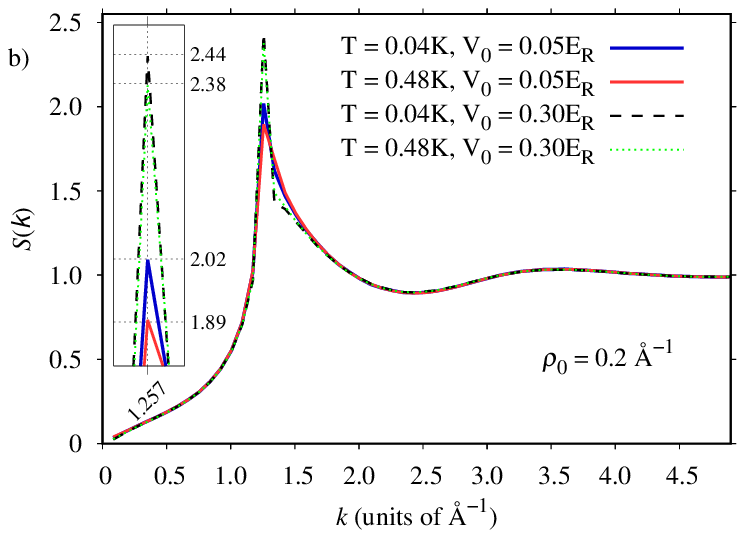}
		\caption{(Color online) Finite-temperature $S(k)$ for two densities (a) $\rho_0$ = 0.1 \AA$^{-1}$ and (b) $\rho_0$ = 0.2 \AA$^{-1}$ for different temperatures and depths. Insets represent zoomed-in peaks of $S(k)$.}
		\label{sofk}
	\end{figure}
	The results for static structure factor $S(k)$ for two densities, depths and temperatures, are presented in Fig. \ref{sofk}. For lower density [Fig. \ref{sofk}(a)], when the depth is a small, there is small difference with respect to the case without optical lattice. With the increase of the temperature, one notices that $S(k)$ starts to increase at small $k$. This appears because $g(x)$ at large distances starts to decay exponentially, which leads to the term proportional to $[(2\pi)^2+(L_Tk)^2]^{-1}$  appearing at small $k$. It is only visible at higher temperature due to smaller $L_T$. At higher depths, for $\rho_0$ = 0.1 \AA$^{-1}$  one additionally notices the appearance of the peak with the wave vector corresponding to the optical lattice potential. A similar peak was observed in the study of bosonic hard rods in a 1D optical lattice~\cite{deSoto:12}. At the density 0.2 \AA$^{-1}$, one observes one peak in $S(k)$ corresponding to $k=2\pi\rho_0$. The peak is enhanced for higher depth of the optical lattice because there is one atom per lattice site. In all cases there is a small temperature dependence. 
	
	\section{Conclusion}
	The low-energy properties of the 1D strongly interacting system with realistic interparticle interactions in a periodic potential with commensurate filling have been determined. The superfluid fraction in the limit of low temperature shows agreement between two used quantum Monte Carlo methods, DMC and PIMC. However, the model from which the dynamical superfluid fraction can be obtained does not fit the data in the superfluid regime well, that is, the best-fit model passes below the PIMC and DMC data at low $LT$. The system was studied thoroughly for four densities, two in the expected robust superfluid phase and two in the fragile superfluid phase. In the robust superfluid phase ($K$ > 2) it was clearly demonstrated, as expected from the LL theory that it takes a finite potential depth of the periodic potential to achieve the transition to Mott insulator. The depth is larger when the Luttinger parameter is larger. Despite the difference in microscopic models, when the bare Luttinger liquid parameters are equal, the obtained values of the critical depth for the superfluid-insulator transition are close to both experimental and theoretical results in ultracold gases~\cite{Haller,Astrakharchik:16,Boeris:16}, demonstrating the LL universality. In the fragile superfluid phase ($K$ < 2), longer simulations and possibly larger lengths are needed to determine with certainty if the extremely small strength of the optical lattice potential destroys the superfluidity, which is expected in the thermodynamic limit. However, as the depth is increased by a small amount, one can clearly observe the loss of superfluidity. Interestingly, when superfluidity is lost the density profile shows that particles are not localized. At small potential depths the correlation functions are not affected. As the depth of the optical lattice is increased one observes oscillations  in the pair-correlation function and the corresponding peak in $S(k)$, while the one-body density matrix demonstrates exponential decay. Exponential decay also leads to the Lorenzian peak in $S(k)$, which can be observed when $L_T$ is not too large. 
	
	It would be interesting to investigate this system at finite temperature for noncommensurate filling of a lattice, in particular focusing on the predicted defect-induced superfluidity~\cite{Astrakharchik:17}.
	
	\section{Acknowledgements}
	
	This research was performed using the resources of computer cluster Isabella based in SRCE, University of Zagreb University Computing Centre and the HYBRID cluster at 
	the University of Split, Faculty of Science and Croatian National
	Grid Infrastructure (CRO NGI) were used. This work was supported by the Croatian Science Foundation under the Project No. IP-2014-09-2452

	
	\bibliographystyle{apsrev}

\end{document}